\documentclass[%
nofootinbib,
 amsmath,amssymb,
 aps,
showkeys,
11pt
]{revtex4-2}

\usepackage{graphicx}
\usepackage{dcolumn}
\usepackage{bm}
\usepackage{subcaption}
\usepackage{amsmath}
\usepackage{hhline}
\usepackage{soul}

\newcommand{\nc}{\textsl{new}cleo}
\newcommand{\SC}{SOURCES}
\newcommand{\SCv}{SOURCES-4C}
\newcommand{\alphan}{($\alpha, n$)}

\newcommand{\CM}{\mathrm{CM}}
\newcommand{\LAB}{\mathrm{LAB}}

\newcommand{\beq}{\begin{equation}}
\newcommand{\eeq}{\end{equation}}

\DeclareUnicodeCharacter{2212}{-}

\begin{document}

\preprint{APS/123-QED}

\title{Investigation of evaluated nuclear data in\\
       the prediction of inherent neutron sources}

\author{Sigtryggur Hauksson}
 \email{Corresponding author: sigtryggur.hauksson@newcleo.com}
\author{Ilaria Casalbore}
\author{Daniele Tomatis}%
\affiliation{%
  Codes \& Methods Dept., \textsl{new}cleo SA,
  7 Bd Gaspard Monge, 91120 Palaiseau, France.
}%

\author{Nunzio Burgio}
\affiliation{NUC-IRAD-RNR, ENEA C.R. Casaccia, Rome Italy.}


\date{\today}

\thanks{We thank Sacha Barré for contributions at early stages of this work.}

\begin{abstract}
Quantifying inherent neutron sources in matter, particularly $(\alpha,n)$ reactions and spontaneous fission, is important in nuclear engineering and other fields. The SOURCES code is a common tool for calculating the yield and spectrum of such neutrons. This paper critically examines all modelling assumptions and nuclear data in SOURCES and proposes alternative approaches where applicable. For $(\alpha,n)$ reactions, we show that the alpha emission lines for $^{235} \mathrm{U}$ should be updated. Furthermore, we compare  four different stopping power data sets for alpha particles slowing down and propose measurements to constrain mixed oxide nuclear fuel data. We use the computer code PHITS to show that energy and angular straggling during the slowing down of alpha particles in the material of interest is unimportant. Then, we compare the cross section and emission spectrum of $(\alpha,n)$ reactions in SOURCES to recently evaluated data libraries. Importantly, the modelling of SOURCES for the emission spectrum seems too simple and may need to be updated. Finally, we compare data on spontaneous fission and show that while the neutron yield from SOURCES is reliable, some discrepancy is found with the neutron spectrum of evaluated data libraries. Complementing this work is an implementation of spontaneous fission in the Monte Carlo code OpenMC.
\end{abstract}


\maketitle


\section{Introduction}
\label{sec:intro}


Neutrons can be emitted spontaneously in certain materials, such as $\mathrm{UO}_2$ and mixed-oxide nuclear fuel (MOX). Careful predictions of these neutrons' yield and emission spectrum are needed to ensure safe operation and disposal of  fresh and spent nuclear fuel. In particular, \nc\ which is currently designing new units of Lead Fast Reactors (LFR) operating with MOX fuel, and which plans to build its own MOX manufacturing plant in France, must carefully quantify inherent sources for radiation protection studies of fuel elements. Such neutron sources in MOX fuel could also be used for the nuclear reactor startup.
Other applications in nuclear engineering that require accurate calculations of spontaneous neutron emission include the enrichment of uranium where $\mathrm{UF}_6$ emits neutrons \cite{Broughton2021}, as well as non-destructive assay measurements. Additionally, fundamental physics experiments such as dark matter searches require the subtraction of spontaneously produced neutrons \cite{CanoOtt_2405,Agnes2018}, and the design of radioisotope thermoelectric generators can require the quantification of inherent neutrons \cite{RTG}. 

The inherent neutron production stems from decaying heavier actinides that usually branch within spontaneous fission and alpha decay. In the latter process, alpha particles are slowed down as they propagate in matter and may hit target light nuclei, such as $^{17} \mathrm{O}$ and $^{18} \mathrm{O}$, inducing neutron production via $(\alpha,n)$ reaction. In addition to these two reactions, delayed neutron emission after fission events can also gives a small contribution. In MOX fuel,  neutrons from $(\alpha,n)$ reactions and spontaneous fission contribute roughly equally. In compounds used in some other applications, such as in $\mathrm{PuF}_4$, $(\alpha,n)$, reactions are dominant.

The computer code \SC\ is usually used to estimate neutron sources in nuclear engineering. \SC\ was originally developed in the early 1980s at LANL \cite{Perry1981,Wilson1983} to calculate neutron production in nuclear oxide fuel. It has undergone a few updates, with the most recent version being  SOURCES-4C, which dates from 2002 \cite{Wilson2009,Wilson2002}. Relative to earlier versions, it includes more recent data for spontaneous fission \cite{Shores2001} and the possibility of a three-region slab problem \cite{Charlton2000}.
In addition to \SC, there are a few other deterministic tools for spontaneous neutron production. These are NEDIS \cite{Vlaskin2017,Bedenko2020}, a Russian computer code that describes the same physics as \SC\ and uses its own evaluated data for $(\alpha,n)$ cross sections \cite{Vlaskin2015, Vlaskin2021}, and neuCBOT
\cite{Westerdale2017,Gromov2023}, which was conceived for dark matter experiments in a homogeneous matter and which does not include spontaneous neutron fission, see also \cite{Mei2009} for a similar approach. Additionally, the Monte Carlo code Geant4 has been used to calculate spontaneous neutron production \cite{Mendoza2020}. In this work, we will focus on \SC, the most widely used tool for nuclear power engineering. All data analyzed comes from SOURCES-4C which is the most recent version of the code. 

Since the most recent update of SOURCES in 2002, new measurements and nuclear data evaluations  suggest re-examining some of the data used by SOURCES, e.g. for the cross section for$(\alpha,n)$ reactions \cite{Perry1981} , which seems to date from over half a century ago. 
In this paper, we carefully examine the modelling assumptions of SOURCES and the data used in the code, comparing it with the most recent data available. The comparison aims to evaluate how including more recent data can improve neutron yield and spectrum predictions and establish whether additional measurements are needed.

This paper is structured as follows. After a brief reminder of the modelling assumptions of SOURCES, Sec. \ref{sec:alphan1} discusses data on alpha lines, and a comparison of stopping power data sets is performed. Besides, we find that the effect of energy and angular straggling is unimportant for the material of interest and the considered alpha energy range. In Sec. \ref{sec:alphan2}, the cross-sections and emission spectra of neutrons in $(\alpha,n)$ reactions are discussed. Finally, in Sec. \ref{sec:SpontaneousFission}, different data about spontaneous fission are compared. Results are summarized in Sec. \ref{sec:conclusions}. Supplementary material on kinematics in $(\alpha,n)$ reactions is contained in App. \ref{App:kin}.

\section{Emission and slowing down of alpha particles}
\label{sec:alphan1}

 SOURCES generally uses a simple model to describe the physics of $(\alpha, n)$ reactions in a homogeneous medium, including alpha/neutrons emitter and target nuclide, employing stopping power data to slow down alpha particles. SOURCES also allows the user to define a more complex set-up: beam problems, where alpha particles from an external beam hit a homogeneous medium, interface problems, where a homogeneous slab of source nuclides is in contact with a homogeneous slab of target nuclides, and three-region interface problems, where a third material is sandwiched between the two slabs of the interface problem. The stopping power $SP(E)=-\frac{dE}{dx}$ is defined as the average energy loss per unit path length of charged particles travelling in matter, and it is one of the main quantities used in the SOURCES simulation. Normalising $SP(E)$ with the atomic density $N$ of the material yields in the stopping cross-section $\epsilon(E)$ ,
\begin{equation}
\label{Eq:StoppingCrossSection}
\epsilon(E) = - \frac{1}{N} \frac{dE}{dx},
\end{equation}
In matter composed of multiple elements, the stopping cross-section of different elements are assumed to add up, so that
\begin{equation}
\label{Eq:Bragg}
\epsilon(E) \approx \frac{1}{N} \sum_j N_j \epsilon_j(E), \quad N=\sum_j N_j,
\end{equation}
where $\epsilon_j$ is the stopping cross-section in matter with only element $j$ of atomic density $N_j$. 
The probability for an $\alpha$ particle with initial energy $E_0$ to induce a reaction while slowing down is then given by
\begin{equation}
\label{Eq:PE0}
\mathbb{P}(E_0) = \sum_i \mathbb{P}_i(E_0) = \sum_i \frac{N_i}{N}\int^{E_0}_0 dE\;  \frac{ \sigma_i(E)}{ \epsilon(E)},
\end{equation}
where the sum is over all target nuclides $i$ and $\sigma_i$ is the \alphan\ cross sections for the specific target nuclide $i$ \cite{Wilson2002}. 
In SOURCES, the rate of neutron production through $(\alpha,n)$ reactions  is then obtained by combining Eq. \eqref{Eq:PE0} (discretized with the trapezoidal rule) and the total rate of alpha particle emission in matter per unit volume as 
\begin{equation}
 \label{Eq:dndt}
\frac{dn}{dt} = \sum_k N_k \lambda_k \sum_l f^{\alpha}_{kl} \mathbb{P}(E^{kl}_0),
\end{equation}
where $\lambda_k$ is the decay constant of nuclide $k$ emitting alpha particles, $f^{\alpha}_{kl}$ is the intensity of alpha line $l$ in that nuclide and $E^{kl}_0$ is the initial energy of the alpha particle from line $l$.
%

In this work, a critical evaluation of data that comes with SOURCES is performed by comparing it with more recent and complete data sets, and by seeing how  results are affected. The data analysed include the intensity of the discrete alpha emission (alpha lines) during decay, the stopping power of elements, the cross-sections for the $(\alpha,n)$ reaction and the resulting neutron source energy distribution. Data on spontaneous fission is treated in section \ref{sec:SpontaneousFission}. We note that an alternative analysis of the data needed for $(\alpha, n)$ reactions was performed in  \cite{Simakov2017}, which focused on yields and did not analyse some of the underlying assumptions in the SOURCES transport coding.
Therefore, in addition to critical analysis of the data sets, we perform an independent review of the SOURCES modelling assumptions for the alpha particles slowing down by comparing them with Monte Carlo simulations performed by the computer code PHITS \cite{phits2002,phits2024}. Further analyses of SOURCES data was performed in  \cite{CanoOtt_2405, Parvu_2408} which focused on the $(\alpha,n)$ cross section. In the meantime, new data evaluations on $(\alpha,n)$ cross section have been performed that are used for comparison with SOURCES data in our study.
%
 
%
%


\subsection{Alpha lines}

The first step in describing $(\alpha,n)$ reactions is the emission of alpha particles. SOURCES contains 619 alpha lines in 96 different nuclides. In SOURCES-4C, no alpha lines beyond 6.5 MeV are included in calculations due to less reliable data on $(\alpha,n)$ cross sections above that energy \cite{Wilson2002}. Alternative versions of the code allow for alpha lines of any energy \cite{Carson2004,Tomasello2008}. For comparison, the ENDF/B-VIII.0 library contains 869 alpha lines for the same 96 nuclides, of which 798 are below 6.5 MeV, evidentiating that the most recent versions of libraries contain substantially more alpha lines than SOURCES. 

In Table. \ref{Table:alphalines}, we compare the number of alpha lines in U and Pu isotopes between SOURCES and ENDF/B-VIII.0 to estimate the reliability of the alpha lines used in MOX and UO$_2$ fuel calculations. For many isotopes, ENDF/B-VIII.0 contains exactly the same alpha lines as SOURCES, and  when it does contain more alpha lines, the total intensity of the alpha lines found in ENDF/B-VIII.0 but not in SOURCES is very small, suggesting that most values used by SOURCES are still reliable. However, there is a very important exception for $^{235}$U. SOURCES contains 11 lines while ENDF/B-VIII.0 contains 21 lines and the extra lines have a total intensity of 6.2\%. Furthermore, the intensity of alpha lines  in SOURCES for $^{235}$U only adds up to 92.7\%. Therefore, SOURCES underestimates the number of alpha particles coming from $^{235} \mathrm{U}$ by about 6 or 7\%. Calculations of 3\% enriched UO$_2$ fuel, will lead to an underestimate of about 1.2\% in the number of alpha particles produced. Therefore, the information on alpha lines in $^{235}$U should be updated. 

\begin{table}[ht]
\centering
\caption{Comparison of the number of alpha lines in SOURCES and ENDF/B-VIII.0 for isotopes of U and Pu. The total intensity of extra alpha lines for each isotope is also shown, i.e. lines that are present in ENDF/B-VIII.0 and not in SOURCES, as a fraction of the total intensity.\label{Table:alphalines}}
\begin{tabular}{| c | c | c | c |}
\hline
  & SOURCES & ENDF/B-VIII.0 & Total intensity of extra lines \\ \hhline{|=|=|=|=|} 
  U232 &  9 & 9 & 0.0   \\ \hline 
  U233 &  30 & 29 &  0.0 \\ \hline 
  U234 &  6 & 6 &  0.0 \\ \hline 
  U235 & 11 & 21  & $6.2 \cdot 10^{-2}$ \\ \hline
  U236 & 3 & 4  & $5.0 \cdot 10^{-7}$ \\ \hline
 U238 &  3 & 3 & 0.0  \\ \hline
  Pu235 & 1  & 0 & 0.0  \\ \hline
 Pu236 &  6 & 13 & $5.0 \cdot 10^{-7}$  \\ \hline
  Pu237 &  10 & 9 & 0.0  \\ \hline
 Pu238 & 14 & 14 & 0.0 \\ \hline
 Pu239 & 3 & 52 & $1.8 \cdot 10^{-3}$   \\ \hline
 Pu240 &  7 & 11 & $1.0 \cdot 10^{-9}$  \\ \hline
 Pu241 & 11 & 11 & 0.0   \\ \hline
 Pu242 & 4 & 4 & 0.0   \\  \hline
 Pu244 & 2 &2 & 0.0  \\ \hline
\end{tabular}
\end{table}

\subsection{Comparison of stopping power data}

Alpha particles lose energy when traversing matter due to Coulomb interactions with atomic nuclei and electrons before potentially inducing $(\alpha,n)$ reactions. As explained above, this slowing down of alpha particles is described with stopping powers in SOURCES which are solely a function of the alpha particle energy. The stopping power can be divided into two contributions: the electronic-stopping power, due to interactions with electrons, and the nuclear-stopping power, due to collisions with nuclei. The first dominates for energetic ions, while the latter prevails at energies of few keV.


As a first step in evaluating the stopping power modeling in SOURCES, we compared its data with those of three other programs that tabulate alpha particle stopping power, namely ASTAR, ATIMA and SRIM. ASTAR is a program based on the ICRU\footnote{International Commission on Radiation Units \& Measurements} report 49 \cite{ICRU49} and has an online interface\footnote{\url{https://physics.nist.gov/PhysRefData/Star/Text/ASTAR.html}} allowing for the calculation of stopping power in 74 elements and compound materials. ATIMA is a program developed at GSI Helmholtz Centre for Heavy Ion Research, which also has an online interface\footnote{\url{https://web-docs.gsi.de/\texttildelow weick/atima/}}. Finally, SRIM consists of a program suite that calculate the stopping and range of ions in matter for elements with Z $\leq 92$ and their compounds. 
All of these databases include both the electronic and nuclear stopping power. Furthermore, SRIM and ATIMA allow for calculations of other quantities, such as the range of alpha particles in matter or the straggling.

The modelling of the stopping power datasets that we compare is based on the same physical principles. High-energy alpha particles were treated with theoretical calculations based on the original works by \cite{Bohr1913, Bethe1930, Bethe1932,Bloch1933,Bohr1948}, where energy loss due to the interaction of the projectile particle with an electron is calculated, considering quantum and relativistic corrections, (see \cite{Lindhard1996, Salvat2022} for further details). The codes include different corrections to these basic results, including the effect of the finite size of the nucleus \cite{Lindhard1996}, spatial variations in the electric field of the projectile (Barkas term) \cite{Jackson1972, Ashley1972, Lindhard1976}, shell corrections due to the quantum distribution of electrons within an atom \cite{Salvat2022}, and the Fermi-density effect of polarization due to the projectile electric field \cite{Sternheimer1971}.  All the codes use fitting formulas based on experimental stopping power data at low energy. In the intermediate energy range, all the codes interpolate between theoretical calculations at high energy and experimental results or models at low energy. The nuclear stopping power is obtained by classical-mechanic orbit calculations \cite{PhysRev.99.1287}. The temperature and density conditions at which stopping power evaluations are conducted are similar in all the codes. There are negligible differences in the stopping power of different isotopes of the same element, provided that the atomic density remains the same.


Despite these similarities, the different codes have some important differences, especially in the energy range used for theoretical results and experimental fits. For the case of alpha particles, ATIMA uses theoretically calculated values roughly above  120 MeV, SRIM above 0.8 MeV and ASTAR above 2 MeV. For energies below 1 MeV, ASTAR uses, for most elements, a fitting formula of Varelas and Biersack  \cite{VARELAS1970213} with numerical coefficients from Ziegler \cite{ziegler1977helium} or Watt \cite{Watt1988}.
In SRIM, the stopping power is considered proportional to the ion energy for energies below one keV.  ATIMA uses an older version of SRIM at energies below 40 MeV. 
\SCv\ contains a list of coefficients to calculate stopping power, evaluated using data by Ziegler et al. \cite{ziegler1977helium} for all Z $\leq 92$ and stopping power coefficients calculated by Perry and Wilson \cite{Perry1981} for $92 < Z \leq 105$.

\begin{figure*}[bh!]
    \centering
    \begin{subfigure}[t]{0.5\textwidth}
        \centering
        \includegraphics[scale=0.33]{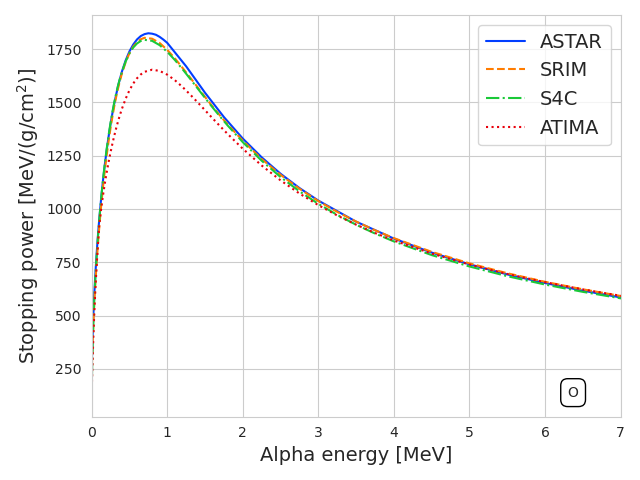}
    \end{subfigure}%
    ~ 
    \begin{subfigure}[t]{0.5\textwidth}
        \centering
        \includegraphics[scale=0.33]{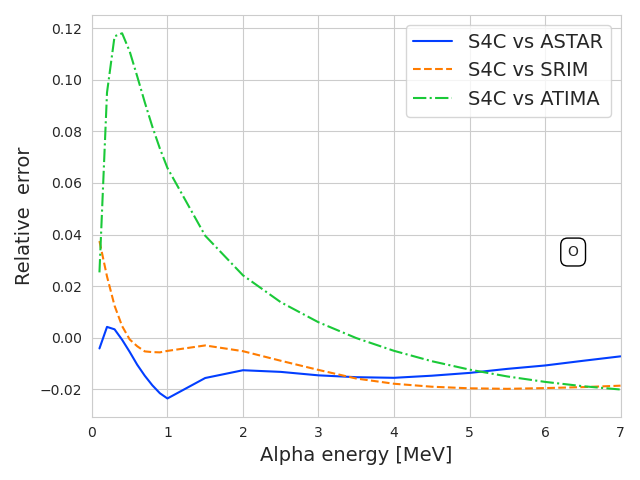}
    \end{subfigure}
	\bigskip
    \begin{subfigure}[t]{0.5\textwidth}
        \centering
        \includegraphics[scale=0.33]{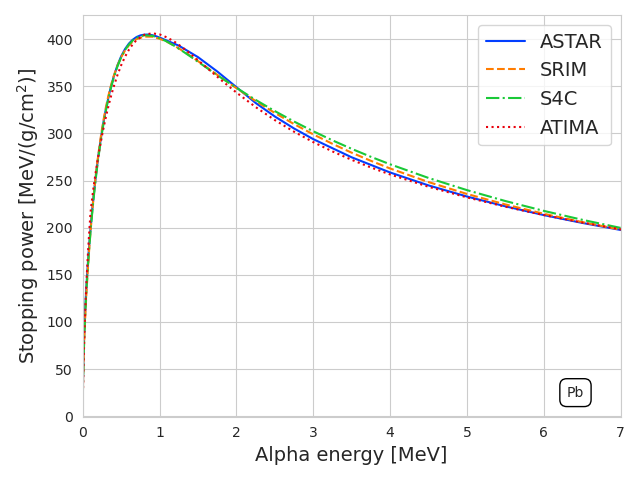}
    \end{subfigure}%
    ~ 
    \begin{subfigure}[t]{0.5\textwidth}
        \centering
        \includegraphics[scale=0.33]{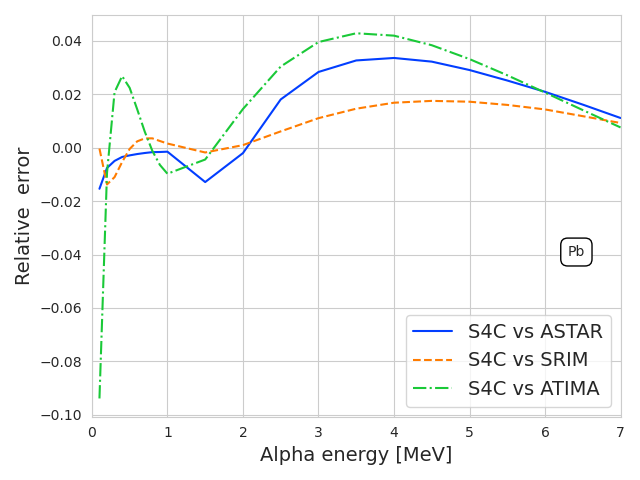}
    \end{subfigure}
	\bigskip
    \begin{subfigure}[t]{0.5\textwidth}
        \centering
        \includegraphics[scale=0.33]{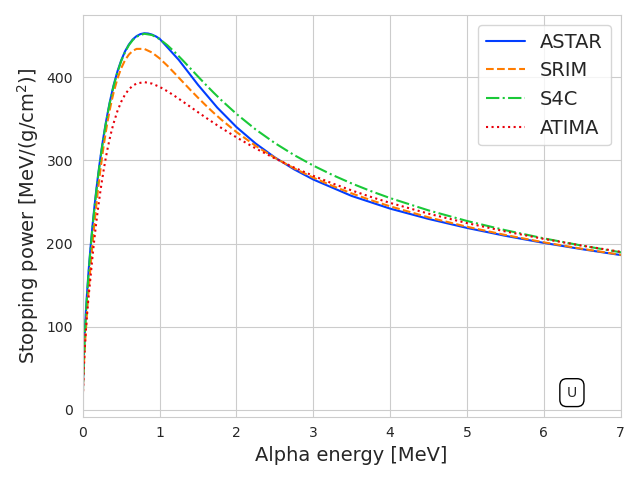}
    \end{subfigure}%
    ~ 
    \begin{subfigure}[t]{0.5\textwidth}
        \centering
        \includegraphics[scale=0.33]{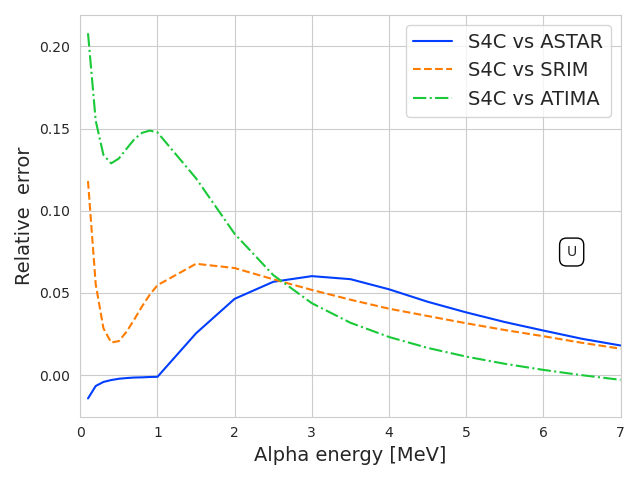}
    \end{subfigure}
    	\bigskip
    \begin{subfigure}[t]{0.5\textwidth}
        \centering
        \includegraphics[scale=0.33]{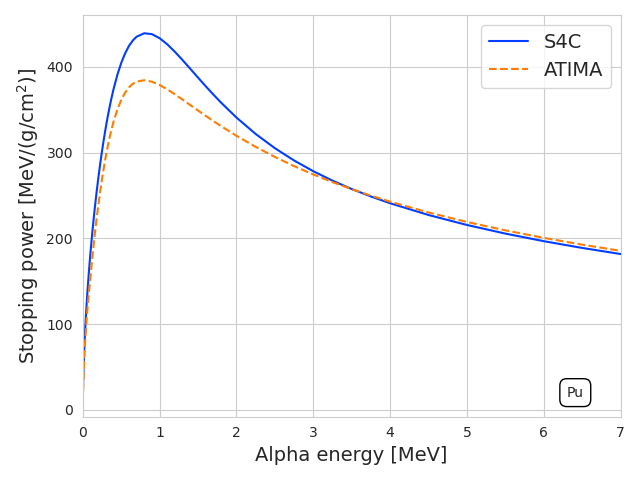}
    \end{subfigure}%
    ~ 
    \begin{subfigure}[t]{0.5\textwidth}
        \centering
        \includegraphics[scale=0.33]{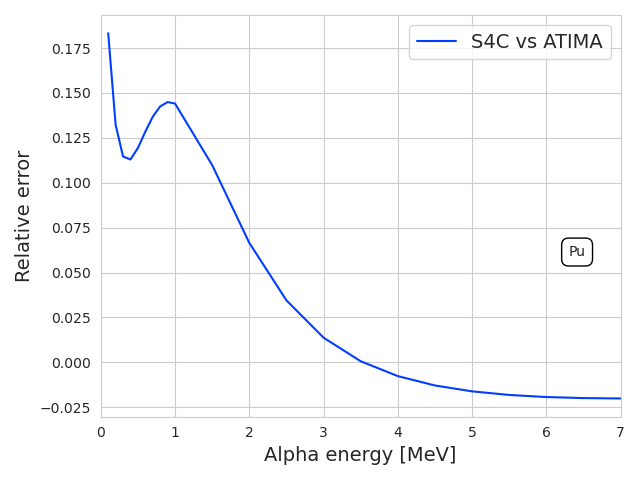}
    \end{subfigure}
    \caption{Comparison of stopping power in 
   O, Pb, U and Pu
    from four different datasets: SOURCES libraries, SRIM, ASTAR and ATIMA. The stopping power in these databases has negligible dependence on the isotope composition. It depends on the atomic density. }
     \label{Fig:stoppow}
\end{figure*}

Fig. \ref{Fig:stoppow} compares \SCv\ stopping powers with the ones of aforementioned codes for Pu, U, O (present in MOX fuel) and Pb.  \SCv{}, SRIM,  and ASTAR mostly show a good agreement. Systematically, ATIMA exhibits larger discrepancies, in particular at low energy.  The significant difference is probably due to the  ATIMA's use of an older set of SRIM stopping powers and the significantly different energy thresholds for theoretical evaluations or models compared to the other codes. The comparison suggests that SRIM, ASTAR and SOURCES stopping power data are suitable for calculating $(\alpha,n)$ reactions. ASTAR is in the public domain and can be freely used. We note that a comparison of stopping power data sets was performed for a select number of elements in e.g. \cite{Kumar2018,Hansson2020} but did not include the ATIMA data set. 
 

\begin{figure}[bh!]
    \centering
        \includegraphics[scale=0.4]{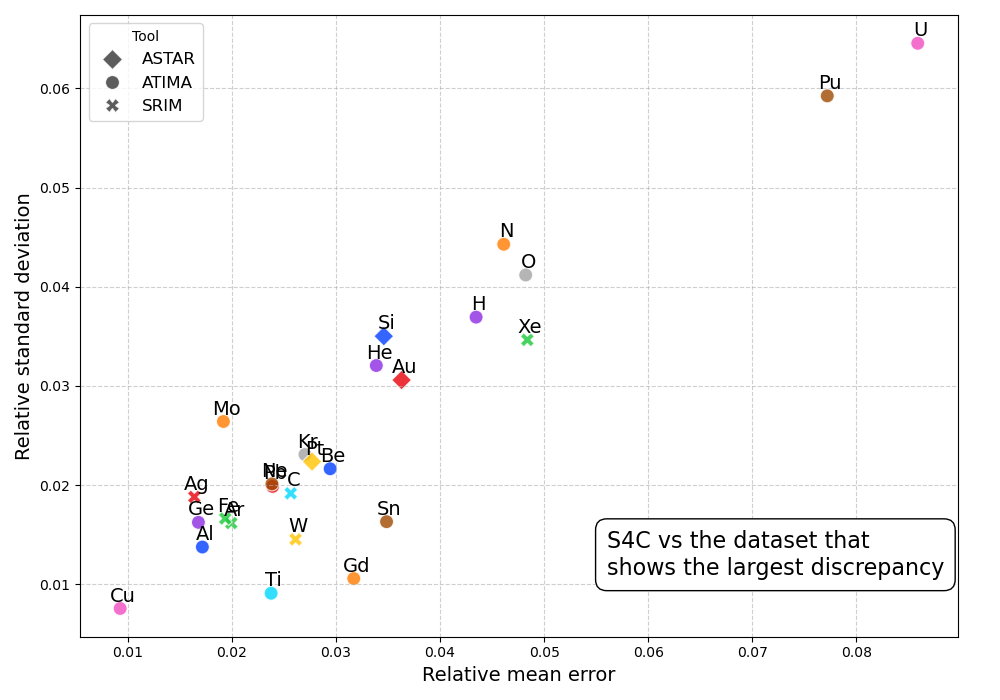}
    \caption{Scatter plot of the relative mean error and relative standard deviation in stopping power between SOURCES and the compared data sets. For each element, only the largest relative mean error between SOURCES and the other three stopping power datasets is shown.}
    \label{Fig:scatter}
\end{figure}

To better estimate the discrepancies between the stopping power datasets under examination, we show in Fig. \ref{Fig:scatter} the relative mean errors and standard deviations, calculated up to 7 MeV, between \SCv\ dataset and the other datasets for various elements of interest. For each element, only the code exhibiting the most significant discrepancy relative to \SCv\ is shown in Fig. \ref{Fig:scatter}, which is most often ATIMA.  Being U, Pu, and O included in the elements with greater uncertainty, the stopping power in MOX fuel, when calculated using  Eq. \eqref{Eq:Bragg}, is equally uncertain. Conversely, datasets using results of stopping power measured directly in MOX fuel should be more accurate. Further measurements might be in order, especially for Pu, which has poorly known stopping power data.
%

\subsection{Straggling in the slowing down of alpha particles with PHITS}

In the stopping power model used in SOURCES, all alpha particles with a given initial energy lose the same amount of energy while traversing the same length.  In reality, each particle can interact in different ways with matter, due to quantum effects and due to differing impact parameter in collisions. This means that particles lose a variable amount of energy when traversing the same length of matter (energy fluctuations), and do not necessarily travel in straight lines. To establish how important this effect is when performing inherent neutron calculations, we have compared results for the slowing down of alpha particles using a stopping power model with results from the Monte Carlo  computer code PHITS \cite{phits2002,phits2024}.

PHITS includes stochastically some degrees of freedom in the slowing down of alpha particles by introducing energy straggling, which accounts for statistical fluctuations in energy loss, $\Delta E$, and angular straggling, which accounts for deviations of the trajectory from a straight line.
The statistical distribution for fluctuations in energy loss comes from the Landau-Vavilov theory \cite{GEANT, Vavilov1957}. Assuming that the alpha particles are non-relativistic with velocity $\beta = v/c$ and that they undergo multiple collisions, the Landau-Vavilov distribution reduces to a Gaussian distribution for energy loss
with standard deviation $\sigma_{\mathrm{energy}} = \sqrt{\xi E_{\mathrm{max}}}$ and a mean energy loss given by the stopping power model. Here $E_{\mathrm{max}}$ is the maximum energy transferred in a collision, and $\xi \approx 614 Z\rho\delta z /A \beta^2$ with $\rho$ the density of the material and $Z$ and $A$ the atomic and mass numbers of the nuclides. Furthermore, $\delta z$ is the length of matter traversed.
%
Angular straggling is modelled similarly with a standard deviation $
\sigma_{\mathrm{angular}} = \sqrt{\langle \Delta \theta^2 \rangle} ,
$
 taken from \cite{LYNCH19916}.

In practice, PHITS calculates the slowing down of alpha particles by splitting up the particle's trajectory in step sizes determined by the parameters \texttt{delt0} and \texttt{deltc}. For each step, mean energy loss is calculated with stopping powers, and energy and angle fluctuations are added by sampling the distributions for energy and angle straggling. The step sizes need to be sufficiently small so that results do not depend on their value.  Default stopping power values in PHITS are taken from the ATIMA data set, but the code supports additional datasets, which can also be implemented by the user. Despite the limitations of the ATIMA data set, it will be used in our analysis of straggling as the results do not depend on the exact stopping power data set chosen.  

We simulated alpha particles hitting a thin layer of matter using PHITS to establish the importance of straggling. The film of matter was a cylinder with thickness $\Delta z$ and a large radius $R$ taken to be much larger than the range of alpha particles in matter. The alpha particles beam was mono-energetic and normally incident to the large surface of the cylinder, with a homogeneous spatial distribution across the surface.  For calculations of average energy loss, we used the T-Deposit tally that estimates the total energy loss of alpha particles in the film volume. We considered all the energy loss deposited locally, turning off the transport of all the secondary particles, avoiding their escaping from the film region.\footnote{For this reason, the T-Let tally for the linear energy transfer gives identical results.} We used the track-length estimator T-track to estimate the fluence particle's angular and energy distributions.

We performed various checks to establish the robustness of this setup and to validate the PHITS model. Firstly, we verified that when straggling is turned off and the thickness $\Delta z$ is small, this setup reproduces the stopping power given by ATIMA. Specifically, we have shown that for thin films, where $\Delta z$ is one percent or less of the range of the alpha particles, $\Delta E / \Delta z$ agrees with the stopping power $dE/dz$ given by ATIMA, with discrepancy of less than $0.15\,\%$, and substantially less at higher alpha particle energy.
We verified that the calculation results were independent of the geometric set-up chosen. Specifically, we defined a simulation set in which we changed the radius of the cylinder over orders of magnitude and used a point-like beam instead of a spatially homogeneous beam normal to the surface of the cylinder. Finally, we tried using a sphere with radius $\Delta z$ and a point-like isotropic source at the centre. In all cases, we get identical results when straggling is turned off. The final test was performed by changing \texttt{delt0} and \texttt{deltc} to obtain different step sizes until the convergence of the results is obtained.  

 \begin{figure*}[b]
    \centering
    \begin{subfigure}[t]{0.5\textwidth}
        \centering
        \includegraphics[scale=0.5]{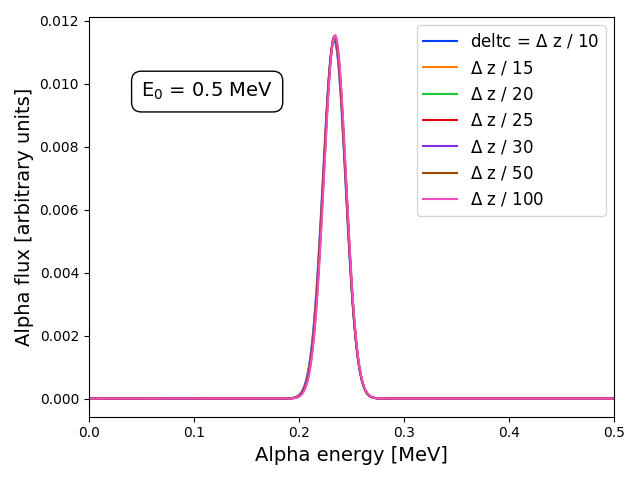}
    \end{subfigure}%
    ~ 
    \begin{subfigure}[t]{0.5\textwidth}
        \centering
        \includegraphics[scale=0.5]{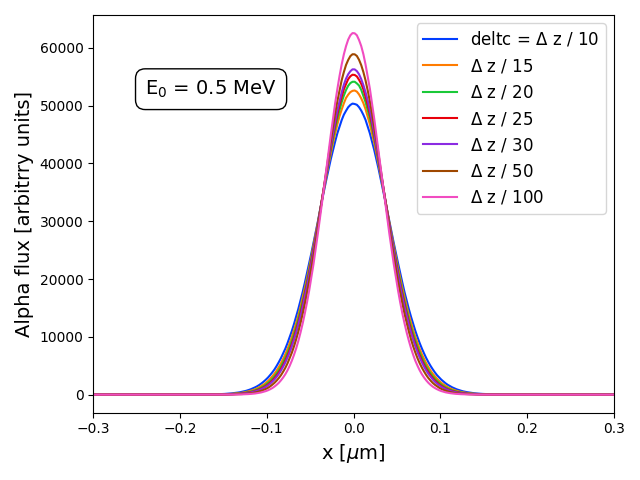}
    \end{subfigure}
    \caption{Distribution of alpha particles that traverse a material of thickness  $0.5\times r(E_0)$ and which have initial energy 0.5 MeV. The left panel shows distribution in energy, while the right panel shows distribution in position transverse to the beam axis. The different curves have varying values of the step size parameter \texttt{deltc}.}
     \label{Fig:alphadistr}
\end{figure*}

\begin{figure*}[b]
    \centering
    \begin{subfigure}[t]{0.5\textwidth}
        \centering
        \includegraphics[scale=0.5]{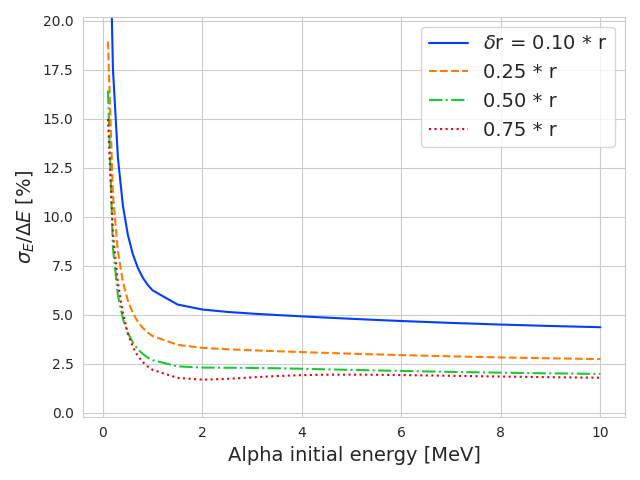}
    \end{subfigure}%
    ~ 
    \begin{subfigure}[t]{0.5\textwidth}
        \centering
        \includegraphics[scale=0.5]{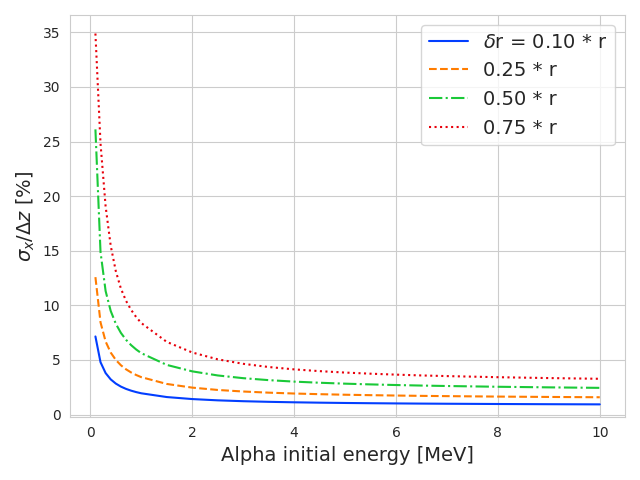}
    \end{subfigure}
    \caption{Relative standard deviation in energy loss (left panel) and transverse position of alpha particles (right panel) traversing matter. The standard deviation is plotted as a function of the initial alpha particle energy, for varying thickness of the traversed material, $\delta r$ with $r$ the particles' range.}
     \label{Fig:alphastd}
\end{figure*}

Using the set-up of alpha particles hitting a film of matter, we could establish the importance of energy and angular straggling. We performed calculations with a wide range of energies of the alpha particles and for differing film thicknesses, $\Delta z$. The thickness was chosen to be a fraction of the range, i.e. $\Delta z = \delta \times r(E_0)$ where $0 \leq \delta \leq 1$ is a fraction independent of energy and $r(E_0)$ is the range (in $\mu{m}$) of an alpha particle with initial energy $E_0$, given by SRIM. For example, Fig. \ref{Fig:alphadistr} shows the energy and angle distribution of alpha particles of initial energy $E_0=0.5$ MeV  after traversing the film with $\delta = 0.5 $. The straggling is considerable and independent of $\texttt{deltc}$ for energy while the angular straggling has a mild dependence on $\texttt{deltc}$. For further simulations we chose $\texttt{deltc} = \Delta z / 10$. To explore the effect of straggling further, we extracted the standard deviation of distributions, such as those in Fig. \ref{Fig:alphadistr}.
 In Fig. \ref{Fig:alphastd}, we show the standard deviation in energy loss $\sigma_{E}$ as a fraction of the average energy loss $\Delta E$, as well as the standard deviation in the transverse position $\sigma_{x}$ as a fraction of the length traversed $\Delta z$ to quantify angular straggling. These quantities are given as a function of the initial energy $E_0$ and for varying lengths of the material. 

The results in Fig. \ref{Fig:alphastd} show that straggling is a moderate effect but that can become important for high precision calculations in complicated geometry, necessitating the use of Monte Carlo codes like PHITS. In the energy interval of interest for inherent neutron production ($E_0$ between 2-6 MeV), the energy straggling as a fraction of total energy loss ranges from 2-5 \% depending on the thickness of the material. It is nearly independent of the initial energy and its value is greater for a smaller thickness. In a homogeneous medium, these straggling effects are unimportant since all alpha particles eventually come to rest and the exact distance they have traversed is not relevant to the neutron yield. However, in more complicated geometries, such as the interface or 3-region interface problem in SOURCES, the energy that an alpha particles carries when entering the target material will vary due to straggling, and thus the $(\alpha,n)$ cross section covered will vary. Thus, energy straggling could modify the neutron yield by a few percent. Similarly, the angular straggling in Fig.  \ref{Fig:alphastd} is between 1-4 \% for the energy range of interest, but it is greater for a larger thickness of the material. However, angular straggling is in general less important, except in very complicated geometries where particles can enter different materials depending on the trajectory they take. We emphasize that effects of energy and angular straggling can only be captured by Monte Carlo codes like PHITS. 


\section{Cross section and emission spectrum of \alphan\ reactions}
\label{sec:alphan2}

\subsection{$(\alpha,n)$ cross section}

The $(\alpha,n)$ cross section is another central ingredient in calculations of the yield of alpha-induced neutrons. SOURCES uses measured data or nuclear model calculations obtained with GNASH  \cite{GNASH1992} for a total of 19 nuclides.\footnote{The list of available isotopes is $^7\mathrm{Li} $, $^9 \mathrm{Be}$, $^{10,11}\mathrm{ B}$, $^{13} \mathrm{C}$, $^{14} \mathrm{N}$, $^{17,18} \mathrm{O}$, $^{19} \mathrm{F}$, $^{21,22} \mathrm{Ne}$, $^{23} \mathrm{Na}$, $^{25,26}\mathrm{ Mg}$, $^{27} \mathrm{Al}$, $^{29,30} \mathrm{Si}$, $^{31} \mathrm{P}$, $^{37} \mathrm{Cl}$.} The measured data and GNASH calculations are both included for four of the listed nuclides. The maximum energy of the alpha particle allowed for the different target nuclides ranges from 6.5 to  11.5 MeV.

We have compared the data in SOURCES with results in the Japanese Evaluated Nuclear Data Library (JENDL) \cite{JENDL5}, which contains 18 nuclides.\footnote{The list of available isotopes is $^{6,7}\mathrm{ Li}$, $^{9}\mathrm{ Be}$, $^{10,11}\mathrm{ B}$, $^{12,13}\mathrm{ C}$, $^{14,15}\mathrm{ N}$, $^{16,17,18}\mathrm{ O}$, $^{19}\mathrm{F}$, $^{23}\mathrm{ Na}$, $^{27}\mathrm{Al}$, $^{28,29,30}\mathrm{ Si}$.} 13 of those nuclides are also found in SOURCES. In addition to JENDL, we have also compared SOURCES data with model calculations performed with the TALYS code system \cite{Koning2012} and compiled in the TALYS Evaluated Nuclear Data Library (TENDL-2019) \cite{Koning2019}. These model calculations are available for around 2800 nuclides. Furthermore, we have compared with data in the recent  ENDF/B-VIII.1 library \cite{Nobre2024}, which includes information on alpha-induced neutron spectrum for five nuclei.\footnote{These are $^{4}\mathrm{ He}$, $^{6}\mathrm{ Li}$, $^{9}\mathrm{ Be}$, $^{17,18}\mathrm{ O}$.} A similar comparison can be found in \cite{CanoOtt_2405, Parvu_2408} but those papers do not include results from the recent ENDF/B-VIII.1 library.

The JENDL-5 and ENDF/B-VIII.1 libraries list the partial cross section for $(\alpha,Xn)$ processes for reaction types with different final-state particles $X$ and different states of the recoil nucleus. These different reaction types are referred to by MT numbers which are defined in \cite{osti_2007538}. For reaction types with only a neutron and the recoil nucleus in the final state, the partial cross sections are given by MT=50,51,52,\ldots, which correspond to the ground state and excited states of the recoil nucleus, or by MT=91 which corresponds to the production of a neutron in the continuum. The sum of these reaction types is given in MT=4, which refers to the sum of all $(\alpha,n)$ processes. Reaction types with more particles in the final state include $(\alpha,2n)$ (MT=16), $(\alpha,n\alpha)$ (MT=22) and $(\alpha,np)$ (MT=28). The total sum of all processes with a neutron in the final state, i.e. all $(\alpha,Xn)$ reaction types, is given by MT=201. In our comparison we have used MT=201, because the quantity we want to evaluate is the total yield of neutrons from all processes.


 \begin{figure*}[b]
    \centering
    \begin{subfigure}[t]{0.5\textwidth}
        \centering
        \includegraphics[scale=0.4]{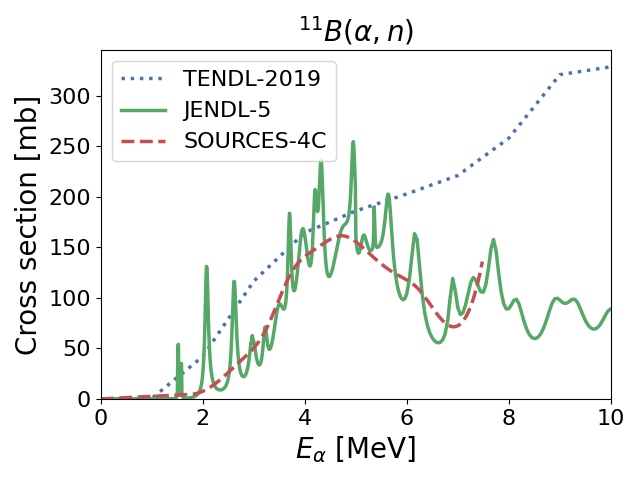}
    \end{subfigure}%
    ~ 
    \begin{subfigure}[t]{0.5\textwidth}
        \centering
        \includegraphics[scale=0.4]{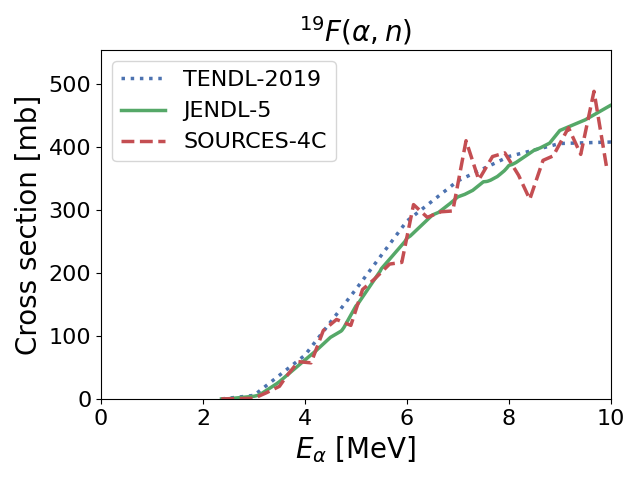}
    \end{subfigure}
	\bigskip
    \begin{subfigure}[t]{0.5\textwidth}
        \centering
        \includegraphics[scale=0.4]{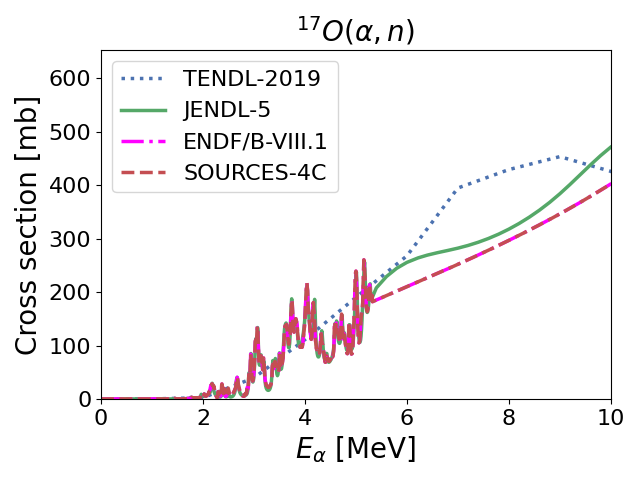}
    \end{subfigure}%
    ~ 
    \begin{subfigure}[t]{0.5\textwidth}
        \centering
        \includegraphics[scale=0.4]{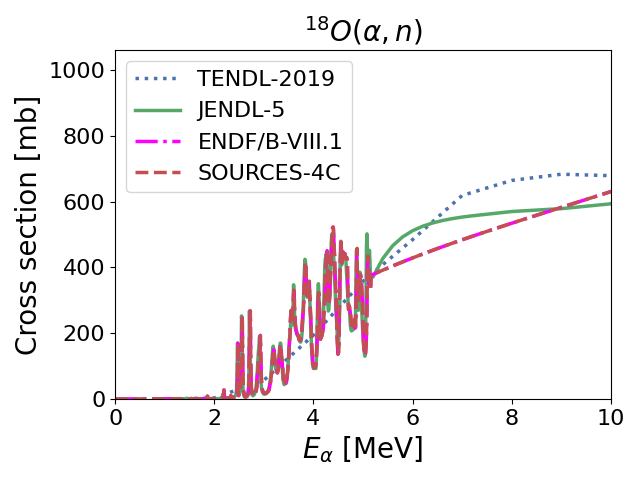}
    \end{subfigure}
    \caption{The cross section of $(\alpha,n)$ reactions as a function of the alpha particle energy. The comparison is performed between Sources4C data sets, JENDL-5 libraries and TENDL-2019 libraries; results are shown for four nuclides.}
     \label{Fig:xsalphan}
\end{figure*}

In Fig. \ref{Fig:xsalphan}, we compare the cross-section of $(\alpha,n)$ reactions for a few selected nuclides. There is good agreement between SOURCES and JENDL-5 below roughly 5 MeV for isotopes of oxygen, while there is a greater discrepancy at higher energy. For the other nuclides, such as isotopes of boron, SOURCES does not report the resonance structure of the cross-section but only the general trend. The TENDL results can only be trusted to estimate the overall shape of the cross section and do not describe the cross-section resonances for any nuclides. All of this suggests that the cross sections in SOURCES could be updated, e.g. using recently evaluated data sets in JENDL-5.

\subsection{Neutron spectrum}

In addition to the total neutron yield, most applications also require the neutron spectrum. In SOURCES, Eq. \eqref{Eq:dndt} for the yield is generalized to give the neutron spectrum\footnote{The neutron spectrum can easily be obtained as $\phi(E_n) = v n(E_n)$ with $v = \sqrt{2E_n/m_n}$ the velocity and $n(E_n)$ the neutron density.} in $(\alpha,n)$ reactions as
\begin{equation}
   \label{Eq:spectrum}
\frac{dn}{dtdE_n} = \sum_k N_k \lambda_k \sum_l f^{\alpha}_{kl}   \sum_i \frac{N_i}{N}\int^{E^{kl}_0}_0 dE_{\alpha}\;  \frac{ \sigma_i(E_{\alpha})}{ \epsilon(E_{\alpha})} \chi(E_n | E_{\alpha}),
\end{equation}
where all quantities are the same as in Eq. \eqref{Eq:dndt} except for $\chi(E_n | E_{\alpha})$, which is the emission spectrum for neutrons with an alpha particle energy $E_{\alpha}$ in the reaction.
 
SOURCES relies on a model to describe the emission spectrum $\chi(E_n | E_{\alpha})$ of $(\alpha,n)$ neutrons, leading to less accurate results. The emission spectrum is given in the laboratory frame, which is defined as the frame of reference where the target nucleus is initially static. The laboratory frame is therefore the frame of the target material when thermal motion of nuclei is ignored.  In the modelling of SOURCES, the alpha particle and the target nucleus form a compound nucleus, which then decays such that the neutron has an isotropic angular distribution in the center-of-mass frame. Using these assumptions, one can show that in the laboratory frame,
\begin{equation}
 \label{Eq:rhotot}
 \chi(E_n| E_{\alpha}) = \sum_m f_m(E_{\alpha}) \rho_m(E_n),
\end{equation}
where $m$ labels different excitation levels of the recoil nucleus, $E_{\mathrm{ex},m}$, and 
\begin{equation}
 \rho_m(E_n) =  \frac{\theta(E_{n,m}^- < E_n < E_{n,m}^+)}{E_{n,m}^+ - E_{n,m}^- } 
\end{equation}
is a uniform distribution with $E_{n,m}^{\pm}$ the maximum and minimum neutron energy given in App. \ref{App:kin}. Here, $\theta(E_{n,m}^- < E_n < E_{n,m}^+)$ is a Heaviside step function guaranteeing that $E_n$ lies between the values $E_{n,m}^-$ and $E_{n,m}^+$, which are determined by the reaction's $Q$-value \cite{Wilson2002}. Furthermore, $f_m(E_{\alpha})$ is the probability for the nucleus to be in a given excitation level, tabulated in SOURCES \cite{Wilson2002}.

We have compared the emission spectrum model of SOURCES with evaluated data in JENDL-5 and ENDF/B-VIII.1 nuclear data libraries. For $(\alpha,n)$ reactions where the recoil nucleus is in the ground state or an excited state (MT=50,51,\ldots), ENDF/B-VIII.1 assumes that the neutron production is isotropic in the center-of-mass frame, just like SOURCES. JENDL-5 goes beyond that by describing the angular distribution in the center-of-mass frame with Legendre polynomials. Furthermore, both JENDL-5 and ENDF/B-VIII.1 include the production of a neutron in the continuum in $(\alpha,n)$ processes (MT=91) as well as processes with more final-state particles, such as  $(\alpha,2n)$ (MT=16), $(\alpha,n\alpha)$ (MT=22) and $(\alpha,np)$ (MT=28), which are parametrized with the Kalbach-Mann representation \cite{KalbachMann1981,Mann1988} which is applicable in the center-of-mass system. We have extracted the energy-angle distributions for the different channels in these two libraries, transformed them to the laboratory frame, integrated out the angular distribution, and summed over the different channels to directly compare with the emission spectrum of SOURCES. Further details can be found in App. \ref{App:kin}. Our results differ from the analysis in \cite{Mendoza2020} in that we use the newly available ENDF/B-VIII.1 data and the most recent JENDL-5 data, which are substantially different from the JENDL/AN-2005 data used in \cite{Mendoza2020} for which important shortcomings have been found in $^{17} \mathrm{O}$, $^{18} \mathrm{O}$  and $^{9} \mathrm{Be}$ \cite{Griesheimer2017}. These shortcomings were due to the absence of angular distributions for reactions where the recoil nucleus is not in the continuum. Furthermore, we include all $(\alpha,Xn)$ reactions, as opposed to only exclusive $(\alpha,n)$ reactions in \cite{Mendoza2020}, as the goal is to calculate total neutron production. Finally, we perform deterministic calculations instead of Monte Carlo calculations.

\begin{figure*}[b]
    \centering
    \begin{subfigure}[t]{0.5\textwidth}
        \centering
        \includegraphics[scale=0.45]{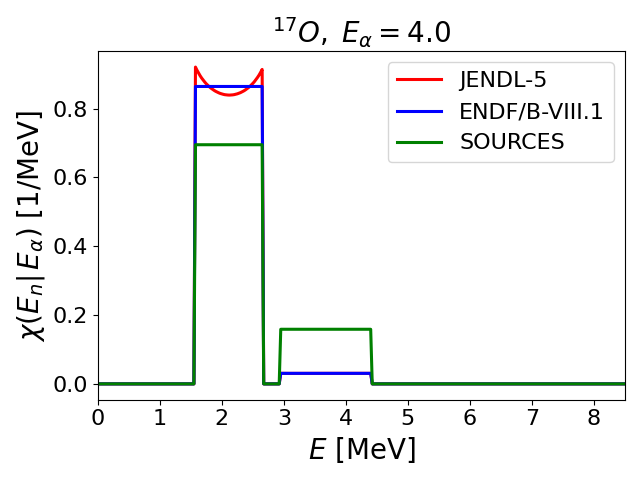}
    \end{subfigure}%
    ~ 
    \begin{subfigure}[t]{0.5\textwidth}
        \centering
        \includegraphics[scale=0.45]{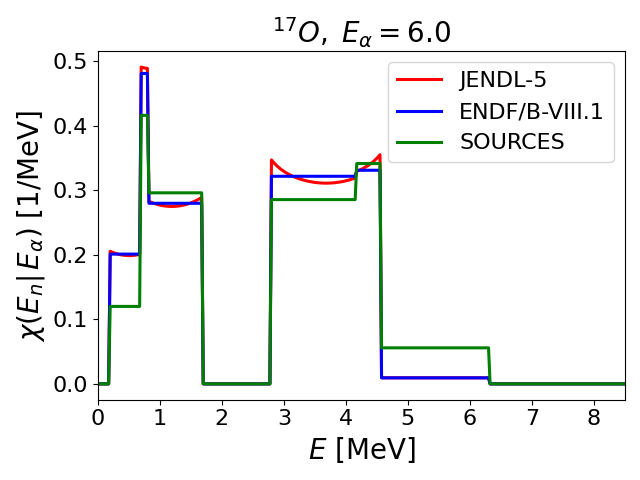}
    \end{subfigure}
	\bigskip
    \begin{subfigure}[t]{0.5\textwidth}
        \centering
        \includegraphics[scale=0.45]{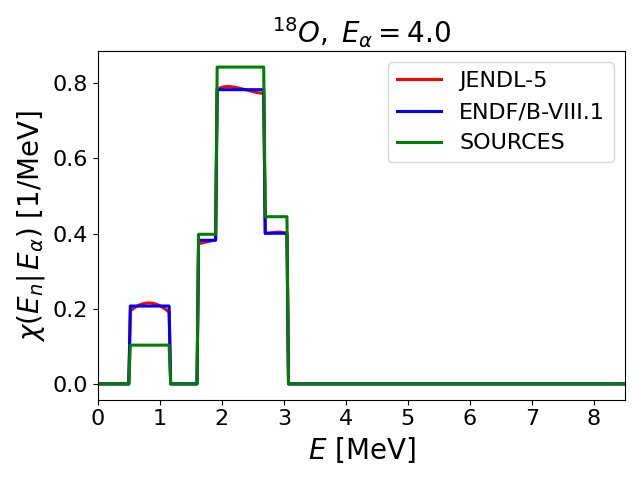}
    \end{subfigure}%
    ~ 
    \begin{subfigure}[t]{0.5\textwidth}
        \centering
        \includegraphics[scale=0.45]{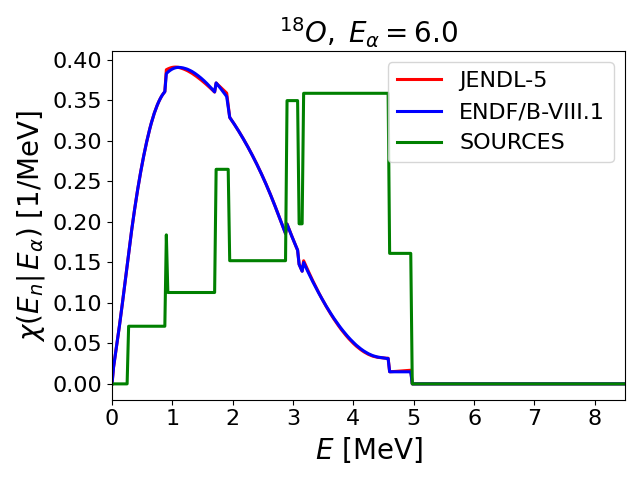}
    \end{subfigure}
    \caption{Distribution of neutron energy in $(\alpha,n)$ reactions, with a $^{17} O$ target and fixed energy of alpha particles. Comparison is performed between modeling in S4C and evaluated nuclear data in JENDL-5 and ENDF/B-VIII.1 libraries.}
     \label{Fig:Endistr}
\end{figure*}

Fig. \ref{Fig:Endistr} compares the neutron energy distribution in the laboratory frame, $\chi(E_n | E_{\alpha})$, for SOURCES, and JENDL-5 and ENDF/B-VIII.1 data, for two values of the alpha particle energy and with both $^{17} \mathrm{O}$ and $^{18} \mathrm{O}$ as targets. At lower energies of the alpha particle, exclusive $(\alpha,n)$ reactions with the recoil nucleus in the ground state or in an excited state (MT=50,51,\ldots) dominate, leading to an emission spectrum that is a sum of uniform distributions, as in Eq. \eqref{Eq:rhotot}. SOURCES correctly captures this behavior but some of the details, notably the probability for the recoil to be in a given excitation level, are different from JENDL-5 and ENDF/B-VIII.1 data. At higher energy, the model of SOURCES completely breaks down because other reactions with a continuous emission spectrum dominate. This is seen already at around $5\,\mathrm{MeV}$ for $^{18} \mathrm{O}$ targets. A similar behavior is seen in the other target nuclei that we have analyzed. This shows that the modeling of SOURCES for the emission spectrum of neutrons is less adequate and more recent evaluations that include all possible reactions must be used in calculations of spontaneous neutrons. This also shows the importance of including as much information as possible about the neutron spectrum in evaluated data libraries. As an example, it would be helpful to include information for more nuclides in the next version of the ENDF/B library.

\begin{figure}[b]
    \centering
        \includegraphics[scale=0.5]{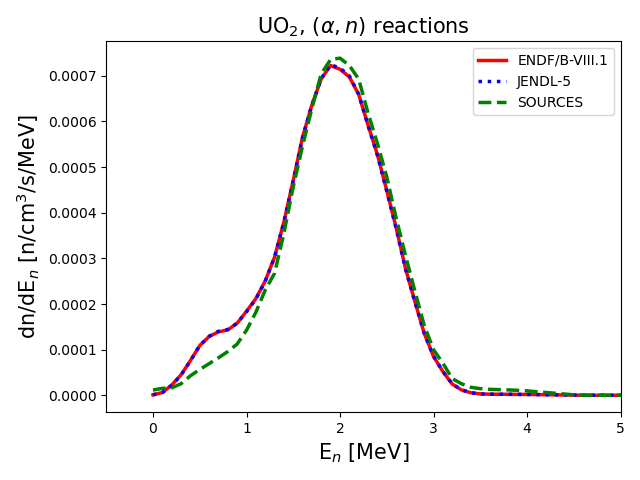}
    \caption{Spectrum of neutrons coming from $(\alpha,n)$ reactions in UO$_2$ fuel. The calculations use the physical modelling of SOURCES, apart from the distribution of neutron energy where one calculation employs the SOURCES modelling while the other one uses experimental data in JENDL or ENDF/B.}
    \label{Fig:spectrum}
\end{figure}

To explore this point further, we have substituted the emission spectrum of JENDL-5 and ENDF/B-VIII.1 in Eq. \eqref{Eq:spectrum} while keeping all other data identical to that in SOURCES, allowing us to isolate the influence of the neutron emission spectrum on the final resulting spectrum. We have performed this calculation for UO$_2$ fuel, see Fig. \ref{Fig:spectrum}. The final results are similar to calculations with the SOURCES emission spectrum, despite the differences seen in Fig. \ref{Fig:Endistr}, because the integration over the alpha particle energy cancels out differences of the emission spectrum at a given alpha energy. We nevertheless recommend using the more detailed emission spectrum of JENDL-5 or ENDF/B-VIII.1 to guarantee the accuracy of the final results. 


\section{Spontaneous fission}
\label{sec:SpontaneousFission}

\subsection{Comparison of spontaneous fission data}

Spontaneous fission is the other main source of inherent neutrons in matter.  SOURCES describes the yield and spectrum of neutrons coming from spontaneous fission with several  parameters. Specifically, the yield of neutrons per unit volume and unit time is
\beq
\frac{d n_{\mathrm{SF}}}{dt} = \sum_i N_i \lambda_i F^{\mathrm{SF}}_i \overline{\nu}_i,
\eeq
where $N_i$ is the atomic density of nuclide $i$, $\lambda_i$ is its decay constant, $F^{\mathrm{SF}}_i$ is the branching fraction, i.e. the probability that a decay event is spontaneous fission (most events are alpha decays), and $\overline{\nu}_i$ is the average number of neutrons released in a spontaneous fission event \cite{Shores2001}. SOURCES includes this information for 41 nuclides (of which two are isomeric states). Of these 41 nuclides, 30 have information about the spectrum of the emitted neutrons.\footnote{The isotopes with spontaneous fission data that include a Watt spectrum are $^{230,232}\mathrm{Th}$, $^{231}\mathrm{Pa}$, $^{233,234,235,236,238}\mathrm{U}$, $^{237}\mathrm{Np}$, $^{236,238,239,240,241,242,244}\mathrm{Pu}$,  $^{241,242\mathrm{m},243}\mathrm{Am}$,  $^{240,242,243,244,245,246,248,250}\mathrm{Cm}$, $^{249}\mathrm{Bk}$, $^{248,252}\mathrm{Cf}$. The isotopes with spontaneous fission data that do not include a Watt spectrum are  $^{249,250,254}\mathrm{Cf}$, $^{253,254,254\mathrm{m},255}\mathrm{Es}$, $^{254,255,256,257}\mathrm{Fm}$.} The Watt distribution gives the emission spectrum according to
\beq
\chi(E) = C^{\prime} e^{-E/a} \sinh \sqrt{b E},
\eeq
where $C^{\prime} = \sqrt{\frac{4}{\pi a^3 b}} e^{-ab/4}$ gives the correct normalization \cite{MCNP}. 

We compare the data in SOURCES with two other data libraries. The first library is contained in a report prepared at Lawrence Livermore National Laboratory (LLNL) \cite{Verbeke2014}, the results of which are used in Monte-Carlo codes \texttt{MCNP}, \texttt{TRIPOLI} and \texttt{Geant}. It lists the average number of neutrons in spontaneous fission and the emission spectrum with a Watt distribution for 18 nuclides\footnote{The list of nuclides is $^{232}\mathrm{Th}$, $^{232,233,234,235,236,238}\mathrm{U}$,$^{237}\mathrm{Np}$, $^{238,239,240,241,242}\mathrm{Pu}$, $^{241}\mathrm{Am}$, $^{242,244}\mathrm{Cm}$, $^{249}\mathrm{Bk}$, $^{252}\mathrm{Cf}$.}. (Branching fractions are not listed.) These nuclides are all in SOURCES, apart from $^{232}$U. The second library we compare with is JEFF-3.3 \cite{Plompen2020}. It is, to the best of our knowledge, the only evaluated library in standard ENDF-6 format that includes information on the spectrum of neutrons in spontaneous fission. It gives the branching fraction and average number of neutrons in fission for 27 nuclides\footnote{The list of nuclides is $^{230,232}\mathrm{Th}$, $^{231}\mathrm{Pa}$, $^{232,234,235,236,238}\mathrm{U}$, $^{236,238,239,240,242,244}\mathrm{Pu}$, $^{241,242\mathrm{m},243}\mathrm{Am}$, $^{242,244,246,248,250}\mathrm{Cm}$, $^{249}\mathrm{Bk}$, $^{249,250,252}\mathrm{Cf}$, $^{253}\mathrm{Es}$.}, which are all present in SOURCES apart from $^{232}$U. The spectrum of neutrons from spontaneous fission is always given for these nuclides. A comparison was also performed with the branching fraction and average neutron number in ENDF/B-VIII.0. Since these values are identical to those of JEFF-3.3 for most nuclides of interest, and since ENDF/B-VIII.0 does not contain the spectrum of emitted neutrons, this comparison is not shown. 


Table \ref{Table:nu} compares $\bar{\nu}$ between SOURCES, JEFF-3.3 and the LLNL report. For isotopes of U and Pu, the discrepancies are always below 1 \%, except for $^{239} \mathrm{Pu}$ for which it is 6.9\%. For a few other nuclides, the discrepancies are substantially higher, e.g. 42 \% for $^{232} \mathrm{Th}$ and 53 \% for $^{230} \mathrm{Th}$ between SOURCES and JEFF-3.3. The picture is similar for the branching fraction, where results from SOURCES and JEFF-3.3 are compared in Table \ref{Table:branching}. For U and Pu isotopes with relatively high branching fractions, the discrepancy is less than 1 \%. In contrast, for other isotopes with lower branching fractions or a lower density in MOX fuel, the discrepancy can be a few per cent, going up to 67 \% for $^{236} \mathrm{Pu}$. Some other nuclides have even higher differences, showing that SOURCES calculates the yield from spontaneous fission with adequate precision for MOX fuel, but for less common nuclides, it is preferable to use JEFF-3.3.
In this work, we do not explicitly compare the decay constant of nuclides as these values are well known and agree very well across different libraries.

\begin{table}[htb]
\centering
\caption{Comparison of the average number of neutrons emitted in spontaneous fission, $\bar{\nu}$, over different libraries.\label{Table:nu}}
\begin{tabular}{| c | c | c |  c | c | c |}
\hline
  & SOURCES  \cite{Wilson2002} & JEFF-3.3  \cite{Plompen2020}  & Verbeke et al.  \cite{Verbeke2014} & max. discrepancy \\ \hhline{|=|=|=|=|=|} 
  U234 &  1.81 & 1.8 &  1.81 & $0.6\,\% $ \\ \hline 
 U235 & 1.86 & 1.87 &  1.86 & $0.5\,\% $ \\ \hline
 U238 &  2.01 & 2.0 &  2.01 & $0.5\,\% $ \\ \hline
 Pu236 & 2.13 & 2.12 &  NA & $0.5\,\% $ \\ \hline
 Pu238 & 2.22 & 2.21 &  2.21& $0.5\,\% $ \\ \hline
 Pu239 & 2.16 & 2.32 &  2.16 & $6.9\,\% $ \\ \hline
 Pu240 &  2.16 & 2.151 &  2.156 & $0.4\,\% $ \\ \hline
 Pu242 & 2.15 & 2.141 &  2.145 &  $0.4\,\% $ \\  \hline
 Pu244 & 2.3 & 2.29 &  NA & $0.4\,\% $ \\ \hhline{|=|=|=|=|=|}
 Th230 & 2.14 & 1.39 &  NA & $53\,\% $ \\ \hline
 Th232 & 2.14 & 1.5 &  2.14 & $42\,\% $\\ \hline
 Cf252  & 3.765  & 3.756 & 3.757 & $0.2\,\% $ \\ \hline
\end{tabular}
\end{table}

\begin{table}[htb]
\centering
\caption{Comparison of the branching fraction for spontaneous fission, $F^{\mathrm{SF}}$ over different libraries.\label{Table:branching}}
\begin{tabular}{| c | c | c | c | c |}
\hline
  & SOURCES \cite{Wilson2002} & JEFF-3.3  \cite{Plompen2020}  & max. discrepancy \\  \hhline{|=|=|=|=|}
  U234 & $1.64\cdot 10^{-11}$ & $1.7\cdot 10^{-11} $  & $3.5  \, \%  $   \\ \hline
 U235 & $7.0 \cdot 10^{-11}$ & $7.2 \cdot 10^{-11}$  & $ 2.8 \, \% $\\ \hline
 U238 &  $5.45 \cdot 10^{-7}$ & $5.46 \cdot 10^{-7}$  & $0.2\, \% $\\ \hline
 Pu236 & $1.37\cdot 10^{-9}$ & $8.2\cdot 10^{-10}$  & $67  \, \%  $   \\ \hline
 Pu238 &  $1.85 \cdot 10^{-9}$ & $1.86 \cdot 10^{-9}$  & $0.5 \, \% $ \\ \hline
 Pu239 & $3.0 \cdot 10^{-12}$ & $3.1 \cdot 10^{-12}$  & $3.3 \, \% $ \\ \hline
 Pu240 & $5.75 \cdot 10^{-8}$ & $5.7 \cdot 10^{-8}$  & $0.9 \, \% $ \\ \hline
 Pu242 & $5.54 \cdot 10^{-6}$ & $5.5 \cdot 10^{-6}$  & $0.7 \, \%  $   \\ \hline
 Pu244 & $0.00121$ & $0.00125$  & $3.2 \, \%  $\\ \hhline{|=|=|=|=|}
Th230 & $3.8\cdot 10^{-14}$ & $2.5\cdot 10^{-13}$  & $85  \, \%  $   \\ \hline
 Th232 & $1.8\cdot 10^{-11}$ & $1.4\cdot 10^{-11}$  & $29  \, \%  $   \\ \hline
 Cf250 & $0.03092$ & $0.00077$  & $3916  \, \%  $   \\ \hline
\end{tabular}
\end{table}

\begin{figure*}[b]
    \centering
    \begin{subfigure}[t]{0.5\textwidth}
        \centering
        \includegraphics[scale=0.4]{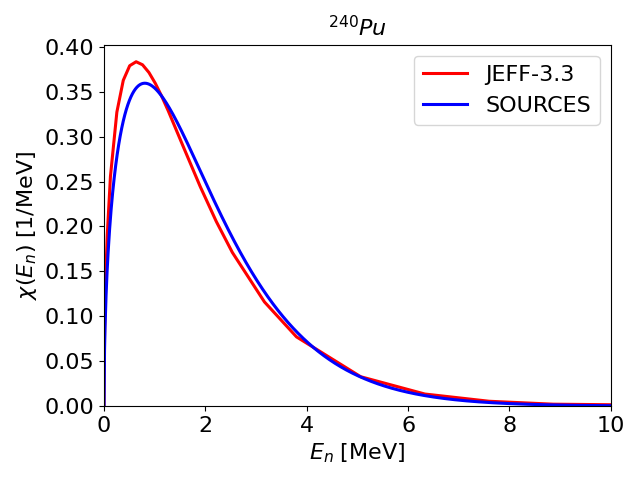}
    \end{subfigure}%
    ~ 
    \begin{subfigure}[t]{0.5\textwidth}
        \centering
        \includegraphics[scale=0.4]{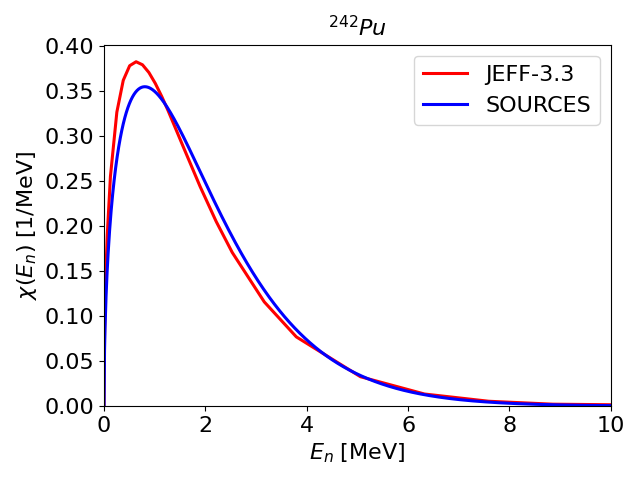}
    \end{subfigure}
	\bigskip
    \begin{subfigure}[t]{0.5\textwidth}
        \centering
        \includegraphics[scale=0.4]{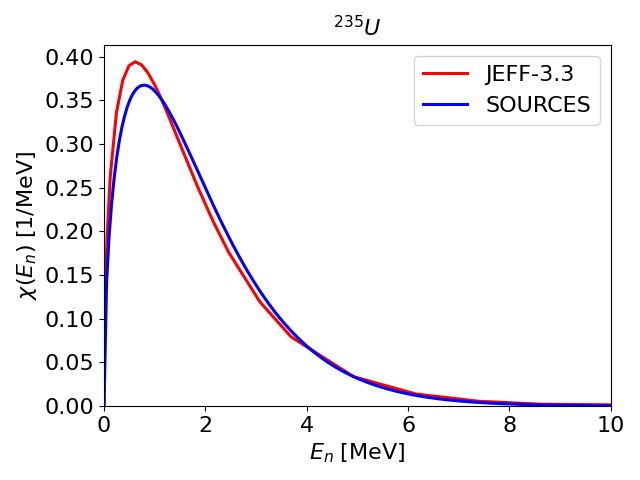}
    \end{subfigure}%
    ~ 
    \begin{subfigure}[t]{0.5\textwidth}
        \centering
        \includegraphics[scale=0.4]{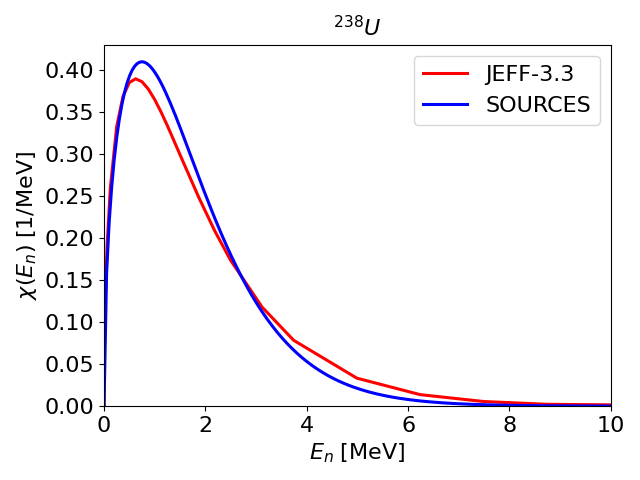}
    \end{subfigure}
    \caption{Comparision of emission spectrum for neutrons in spontaneous fission.}
     \label{Fig:SFspectrum}
\end{figure*}

In Fig. \ref{Fig:SFspectrum}, we compare the emission spectra in SOURCES and JEFF-3.3 for a few isotopes of U and Pu.\footnote{We do not show the parametrization from the LLNL report \cite{Verbeke2014} since it is also a Watt spectrum with virtually identical parameters as in SOURCES.} In general, SOURCES tends to overestimate the energy of the emitted neutrons relative to JEFF-3.3. It furthermore uses a parametrization with a Watt spectrum known to introduce biases, while JEFF-3.3 gives the emission spectrum without any parametrization. For a more precise comparison, we have compared the average energy of emitted neutrons,
\beq
\langle E \rangle = \int dE\; E \,\chi(E)
\eeq
in Table \ref{Table:Ebar}, see also \cite{Endo2005} for a similar analysis. The relative discrepancy varies from a few percent up to 16\% for $^{238} \mathrm{U}$. This higher discrepancy for $^{238} \mathrm{U}$ is important since in $\mathrm{UO}_2$ fuel nearly all neutrons coming from spontaneous fission are due to that isotope.  Furthermore, most neutrons in $\mathrm{PuO}_2$ fuel come from spontaneous fission from $^{240} \mathrm{Pu}$ and $^{242} \mathrm{Pu}$, where the relative discrepancy in the average energy is 4\% and 8\%, respectively.

Our analysis shows that in applications with conventional nuclear fuel, SOURCES results are reliable for yielding neutrons from spontaneous fission. However, further measurements and data evaluation are needed for the spectrum due to the much greater discrepancy in the spectrum when comparing results from SOURCES with JEFF-3.3 (the only evaluated data library available with spontaneous fission). We encourage evaluators of other nuclear data libraries to include spontaneous fission in future evaluations. Furthermore, further measurements of the neutron spectrum in spontaneous fission might be required for some isotopes in MOX fuel, in particular $^{238} \mathrm{U}$, $^{240} \mathrm{Pu}$ and $^{242} \mathrm{Pu}$, which account for most of the yield.

\begin{table}[h!]
\centering
\caption{Comparison of the average energy of neutrons emitted in spontaneous fission over different libraries.\label{Table:Ebar}}
\begin{tabular}{| c | c | c |  c | c |}
\hline
  & SOURCES  \cite{Wilson2002} & JEFF-3.3  \cite{Plompen2020} & max. discrepancy \\ \hhline{|=|=|=|=|} 
  U234 &  1.89 & 1.85 & $2   \, \%  $\\ \hline 
 U235 & 1.83 &1.85 & $-1   \, \%  $\\ \hline
 U238 &  2.24 &1.90 & $16   \, \%  $ \\ \hline
 Pu236 & 2.02 &1.91 & $6  \, \%  $  \\ \hline
 Pu238 &  2.07 &1.93 & $7  \, \%  $ \\ \hline
 Pu239 &  1.93 &1.90 & $2  \, \%  $ \\ \hline
 Pu240 & 1.96 &1.89 & $4  \, \%  $   \\ \hline
 Pu242 &  1.77 &1.92 & $-8  \, \%  $  \\  \hline
 Pu244 & 2.15 &1.96 & $9  \, \%  $ \\ \hhline{|=|=|=|=|=|}
 Th230 & 1.60 &1.76 & $-10  \, \%  $ \\ \hline
 Th232 & 1.92 &1.81 & $6  \, \%  $ \\ \hline
 Cf250  & 2.31 &2.15 & $7  \, \%  $ \\ \hline
\end{tabular}
\end{table}

\subsection{Implementation of spontaneous fission in OpenMC}

Accompanying this work is an implementation\footnote{This development can be found on \url{https://github.com/sigtryggur-hauksson/openmc/tree/develop_sf}. A pull request for merging with the main OpenMC project is on \url{https://github.com/openmc-dev/openmc/pull/3532}.} of spontaneous fission yields in the Monte Carlo code OpenMC \cite{Romano2015}. This open source code performs neutron and photon transport simulations, and can couple transport with depletion calculations that solve the Bateman equation, accounting for gain and loss in nuclide density due to radioactive decay and neutron-induced processes, including neutron-induced fission. Earlier versions of OpenMC could only include spontaneous fission as a loss term (as part of the total decay constant) but did not include gain terms due to the creation of fission products.
Our development introduces such spontaneous fission gain terms when solving the Bateman equation. This can be relevant for applications at low power with nuclides that have a relatively high rate of spontaneous fission. The development furthermore introduces the creation of decay chains from evaluated nuclear data libraries including spontaneous fission yield (sfy) files. Standard operations, such as writing and reading xml files, reducing chains and checking their validity, were extended to account for spontaneous fission. It should be emphasized that spontaneous fission is not included as a source in the Boltzmann equation with this implementation since it is always considered as negligible in reactor calculations with respect to neutron-induced fission multiplication.

\section{Conclusions}
\label{sec:conclusions}

In this paper, we have scrutinized all data and modelling assumptions that are used to evaluate spontaneous neutron production in matter. Such production is predominantly due to spontaneous fission and $(\alpha,n)$ reactions. It finds important applications in e.g. the treatment and transport of nuclear fuel. For concreteness, we analysed the data and modelling in SOURCES, which is a widely used tool for inherent neutron production.  

To calculate neutrons coming from $(\alpha,n)$ reactions, one needs data for the emission of alpha particles, their slowing down in matter, the cross section of the $(\alpha,n)$ reaction, and the neutron emission spectrum. Comparing data in SOURCES with that of recent evaluated data libraries, we showed that alpha emission lines are reliable, except for $^{235} \mathrm{U}$ where discrepancies were found. We furthermore performed a detailed analysis of the slowing down of alpha particles in matter. Comparing four different data sets for the alpha stopping power, we showed that the stopping power in SOURCES is reliable and that the data set ASTAR can be recommended among those that are in the public domain. Nevertheless, further measurements of slowing down of alpha particles in Pu, as well as in MOX fuel would be in order, as the stopping power in these materials is less well known. We furthermore evaluated the importance of straggling, i.e. variation in energy loss and direction, for the slowing down of alpha particles in matter, using the Monte Carlo code PHITS. We found that the inclusion of straggling can be required in high-fidelity calculations, where alpha particles traverse multiple materials. 

The cross section of $(\alpha,n)$ reactions in SOURCES was found to agree well with those in the evaluated data library JENDL for the isotopes of oxygen. However, for some other nuclides, JENDL includes more resonance structure and should be preferred. Analyzing the spectrum of neutrons coming from $(\alpha,n)$ reactions, we showed that simple resonance model of SOURCES breaks down at higher energies of alpha particles, as was seen by comparing with JENDL. Therefore, it is important to represent the neutron spectrum more realistically in calculations. We encourage data evaluators to include the spectrum of neutrons for more isotopes, as this is an essential quantity in inherent neutron production.

Finally, we analyzed neutrons coming from spontaneous fission. For typical nuclides found in MOX fuel, the branching fraction and average number of neutrons emitted agree well between SOURCES and other data libraries, meaning that SOURCES gives the yield accurately. However, there is a greater discrepancy for the spectrum of neutron emitted, especially for $^{238} \mathrm{U}$, between data in SOURCES and JEFF-3.3. We suggest to include the spectrum of spontaneous fission neutrons in more evaluated data libraries. Furthermore, additional measurements for certain isotopes of uranium and plutonium, in particular $^{238} \mathrm{U}$, $^{240} \mathrm{Pu}$ and $^{242} \mathrm{Pu}$ might be required. Complementary to this work of data comparison, we have introduced spontaneous fission in OpenMC, a widely used Monte Carlo neutron transport code.


\section*{Statements and Declarations}

The authors declare no competing interests. 

\section*{Data availability statement}

Most data analyzed in this work is publicly available, with references provided. Other data, such as simulation parameters, are available from the corresponding author on reasonable request.

\appendix

\section{Kinematics in $(\alpha,n)$ reactions}
\label{App:kin}

The nuclear data for the spectrum of neutrons in $(\alpha,n)$ reactions in JENDL-5 and ENDF/B-VIII.1 is given as product energy-angle distributions (MF=6) in the center-of-mass frame. In order to compare this data with SOURCES data, one must transform it to the laboratory frame and integrate out the angle distribution so that only the energy dependence of the neutron is left. A straightforward calculation using energy and momentum conservation shows that the energy-angle distributions in the two frames are related by  
 \beq
\label{Eq:result_transf}
\widetilde{f}(E_{\LAB},\mu_{\LAB};E_{\alpha}) = \sqrt{\frac{E_{\LAB}}{E_{\CM}}} f(E_{\mathrm{CM}},\mu_{\mathrm{CM}};E_{\alpha}).
\eeq
where \(\mu = \cos \theta\) with $\theta$ the angle between the outgoing neutron and the incoming alpha particle and $E_{\LAB}$ and $E_{\CM}$ are the neutron energy in the laboratory and the center-of-mass frame. The energy and the angle in the two frames can be shown to be related by 
\beq
\label{Eq:ECM}
E_{\CM} =  E_{\LAB} - m_n \sqrt{\frac{2 E_{\LAB}}{m_n}} V_{CM}\mu_{\LAB} + \frac{1}{2} m_n V^2_{\CM}
\eeq
and
\beq
\label{Eq:muCM}
\mu_{\CM}   = \frac{\sqrt{2E_{\LAB}/m_n}\,\mu_{\LAB} - V_{\CM}}{\sqrt{ 2E_{\LAB}/m_n - 2\sqrt{\frac{2 E_{\LAB}}{m_n}} V_{CM}\mu_{\LAB} +  V^2_{\CM}}}.
\eeq
Here $V_{\CM} = \frac{m_{\alpha}}{m_t + m_{\alpha}} \sqrt{\frac{2 E_{\alpha}}{m_{\alpha}}}$ is the relative velocity between the two frames, $m_t$ is the mass of the target nucleus and $m_n$ is the mass of the neutron.

We are interested in calculating
\beq
\label{Eq:muint}
\widetilde{f}(E_{\LAB};E_{\alpha}) = \int_{-1}^1 d\mu_{\LAB} \; \widetilde{f}(E_{\LAB},\mu_{\LAB};E_{\alpha}).
\eeq
For the model in SOURCES, the center-of-mass distribution is isotropic in the angle and the neutron can only take one energy value. In other words,
\beq
f\left( E_{\CM},\mu_{CM},E_{\alpha}\right) = \frac{1}{2} \sum_m f_{m}(E_{\alpha}) \, \delta(E_{\CM} - E_{m}). 
\eeq
where the sum over $m$ accounts for different resonance states of the recoil nucleus. Substituting this in Eq. \eqref{Eq:muint} and integrating over the delta function using Eq. \eqref{Eq:ECM} gives that 
\beq
\widetilde{f}(E_{\LAB};E_{\alpha}) = \sum_m f_m \frac{1}{2\sqrt{2m_n E_m} V_{\CM}} \theta\left(E^m_{\LAB,-} < E_{\mathrm{LAB}} < E^m_{\LAB,+}\right)
\eeq 
As is argued in \cite{Wilson2002}, 
\beq
\label{Eq:Emin}
E^m_{\LAB,\pm} = E_m \pm \sqrt{2m_n E_m} V_{\CM}  + \frac{1}{2} m_n V_{\CM}^2.
\eeq
with $E_m = \left( Q - E_{\mathrm{ex}} \right) \frac{m_r}{m_r + m_n}  + E_{\alpha} \frac{m_t}{m_t + m_{\alpha}} \frac{m_r}{m_r + m_n}$, where $m_r$ and $m_{\alpha}$ are the recoil nucleus mass and alpha mass, $Q$ is the $Q$ value of the reaction and $E_{\mathrm{ex}}$ is the excitation energy of the recoil nucleus. 

JENDL-5 relaxes the isotropic assumption for the energy-angle distribution of neutrons for processes where the recoil nucleus is in an excited state (MT=50,51,...). The center-of-mass distribution for a specific excitation level is then of the form 
\beq
f(E_{\CM},\mu_{\CM}; E_{\alpha}) = \delta(E_{\CM} - E_m) g(\mu_{\CM};E_{\alpha})
\eeq 
where $g$ is a probability distribution given by a sum of Legendre polynomials. A similar calculation as above then shows that
\beq
\begin{split}
\widetilde{f}(E_{\LAB};E_{\alpha}) &= \frac{1}{\sqrt{2E_i m_n} V_{CM}}  \theta\left(E^i_{\LAB,\mathrm{min}} < E_{\mathrm{LAB}} < E^i_{\LAB,\mathrm{max}}\right) \\
 &\times g\left(\frac{1}{\sqrt{2m_n E_i} V_{\CM}} \left(E_{\LAB} - E_i - \frac{1}{2}m_n V_{\CM}^2 \right);E_{\alpha} \right)
 \end{split}
\eeq
Finally, in the general case of processes where the recoil nucleus is in the continuum, one needs to perform the integration in Eq. \eqref{Eq:muint} numerically where the center-of-mass distribution in Eq. \eqref{Eq:result_transf} has been extracted from the data libraries. We have performed these calculations individually for all $(\alpha,X n)$ reactions giving an emission distribution $\rho_{\mathrm{MT}}(E | E_{\alpha})$ for every relevant MT number. The total emission spectrum is then given by 
\beq
\rho(E | E_{\alpha}) = \frac{1}{\sigma_{\mathrm{MT}}} \sum_{\mathrm{MT}} \sigma_{\mathrm{MT}}(E_{\alpha}) \rho_{\mathrm{MT}}(E | E_{\alpha})
\eeq
where $\sigma_{\mathrm{MT}}$ is the partial cross section for reaction MT and $\sigma_{\mathrm{MT}} =  \sum_{\mathrm{MT}} \sigma_{MT}(E_{\alpha}) $ is the total cross section.

\bibliography{apssamp}

\end{document}